\documentclass{article}

\usepackage{amsmath,mathrsfs}
\usepackage{ upgreek }
\usepackage{ amssymb }
\usepackage{graphicx,subfigure}
\graphicspath{{v-phi/}}
\usepackage{multirow,booktabs}
\usepackage[numbers,sort&compress]{natbib}
\def\abstract#1{\vskip 7mm 
        \begin{center}{\large Abstract}\par \smallskip
                \begin{minipage}[c]{12cm}
                        \small #1
                \end{minipage}
        \end{center}
}
\def\title#1{\begin{center}{\Large\bf #1}\end{center}}
\def\author#1{\vskip 5mm \begin{center}{#1}\end{center}}
\def\address#1{\begin{center}{\it #1}\end{center}}
\makeatletter
\@ifundefined{lesssim}{}{}
\@ifundefined{gtrsim}{}{}
\def\vereq#1#2{\lower3pt\vbox{\baselineskip1.5pt \lineskip1.5pt
\ialign{$\m@th#1\hfill##\hfil$\crcr#2\crcr\sim\crcr}}}
\makeatother

\begin{document}

\title{	The Scattering of Dirac and Majorana Fermions in Spherically Symmetric Gravitational Field and Torsion Field}
\author{ Junhui Lai$^a$, Xun Xue$^{a,b,}\footnote{Corresponding author:xxue@phy.ecnu.edu.cn}$}
\address{ $^a$Department of Physics, East China Normal University, Shanghai 200241, China}
\address{ $^b$Center for Theoretical Physics, Xinjiang University, Urumqi 830046, China}
\abstract{
	The possibility to distinguish Dirac from Majorana fermion by gravitational interaction and torsion coupling is discussed. The scattering of both Dirac and Majorana particles are the same in the spherically symmetric gravitational field up to the first order perturbation of the fermion-gravity interaction. The scattering behaviors of Dirac and Majorana fermion are also the same by axial vector torsion but different by vector torsion. We get the conclusion that the scattering by vector torsion field can distinguish the Dirac from Majorana neutrino.
	}

\section{Introduction}

Majorana Fermion is first proposed by Majorana in 1937 that the neutral spin one half fermion can be described by the real valued wave function and therefore the fermion is its own antiparticle\cite{Majorana:1937vz}. The idea of Majorana nature of neutrino is proposed by Furry in 1939\cite{Furry:1939qr}, the neutrinoless double beta decay, which only occurs for the Majorana type of neutrino particle. The Standard Model fermions are all known to behave as Dirac particles at the energy scale lower than the electroweak symmetry breaking scale except neutrino, the nature of which is not known. The behavior of neutrino at low energy scale depends on the fermion type of right-handed neutrino introduced to explain neutrino oscillation. However the Majorana nature of neutrino violates the lepton number conservation and even the $B-L$ conservation. Though the neutrinoless double beta decay play an important role in solving the absolute neutrino masses and the neutrino mass hierarchy problems, it has not been discovered yet\cite{Oberauer:2020mdv}. In principle, magnetic field could also distinguish the Dirac or the Majorana type of neutrinos for their unique magnetic moment structures, though the required fields are extremely strong. 

The idea of distinguishing the types of neutrino by the their response to the gravitational effect is proposed since 1990s\cite{Ng:1993vh, Ng:1994qy, Singh:2006ad, Menon:2008wa, Alavi:2012np}. The spin-gravity effect is proposed to distinguish the fermion type of the neutrino by the spin-flip matrix elements induced by the spin-gravity coupling in a Lense-Thirring background which is different for Dirac type of neutrino from Majorana one\cite{Singh:2006ad,Singh:2006fn}. However, the requirement of self charge conjugation of spinor rather than the neutrino field can only apply to the quantum mechanical Majorana type of wave packet rather than a Majorana particle in propagation, which is a plane-wave expansion in quantum field theory approach just as pointed out in reference\cite{Nieves:2006tq}.  The conclusion about the spin-gravity coupling induced spin-flip matrix elements does not follow for the quantum field theory approach and the difference between Majorana and Dirac particle is suppressed by the factor $m/E$ and undetectable in the case\cite{Nieves:2006tq}. 

As a matter of fact, a momentum eigenstate can not be charge conjugation invariant at the same time in a quantum mechanical sense. There is the similar problem in the wave function treatment of neutral scalar particle that there is no momentum eigenfunction which is charge conjugation invariant. The proper treatment framework is the second quantization scheme. The response of neutrino to the gravity is very weak in whatever the direct gravitational vertex coupling or the spin-gravity coupling sense so the effect can only be significant in the astronomical scale propagation of the neutrino. The wave packet approach with different momentum components is not applicable at such large scale propagation. We tend to consider the scattering of asymptotic free neutrino by a weak spherically symmetric gravitational field for both torsionless and torsionness cases in this paper.

The connection is an independent degree of freedom of geometry in general. The Platini formalism of general relativity constrains the connection of spacetime to Levi-Civita one in the absence of spinor matter source. The Einstein-Cartan spacetime is a natural extension of gravitation theory beyond general relativity. The general relativity is usually considered as an effective theory at low energy scale. The connection is constrained or condensed to torsionless Levi-Civita one. At high energy, the connection may be an active degree of freedom and deviates from Levi-Civita one by excitation\cite{Donoghue:2016vck}. Moreover, the non-trivial distribution of spin angular momentum tensor also induces curved spacetime with torsion. Even in the classical gravitation theory which is considered as the low energy effective theory of quantum gravity, torsion can also play an important role. It is revealed that gravity can be described equivalently by either the torsionless curved spacetime with zero nonmetricity or the curvature-less spacetime with torsion and zero nonmetricity or even both torsionless and curvature-less spacetime with dynamical nonmetricity\cite{Arcos:2004ig, Hammond:2002rm}. In the teleparalelle gravity framework, gravity is described by dynamical torsion of the flat spacetime with zero nonmetricity\cite{Aldrovandi:2013wha}. 

Torsion can be decomposed into vector, axial vector and pure tensor parts according to the irreducible representation of Lorentz group\cite{Aldrovandi:2013wha}. The pure tensor part does not appear in the coupling of torsion with fermion and so does the vector part in the minimal coupling scheme. However, the coupling between torsion and fermion is an effective one at low energy scale in the effective theory of high energy gravitational theory such as quantum gravity. The non-minimal coupling between vector torsion and fermion is possible in general\cite{Shapiro:2001rz}. Moreover, in the equivalent description of gravitation by torsion rather than curvature, the vector torsion coupling with fermion does appear. We take both vector torsion and axial vector torsion into consideration in this paper. 

The torsion free Levi-Civita connection in general relativity is the consequence of both the absence of spinor source matter and the local Lorentz invariance. Torsion naturally appears in a gravitation theory with Lorentz violation even in the absence of spinor matter source\cite{wu2015effective, wu2016sim, yang2017VSR, wei2017E2}. It is proposed that the Lorentz violation of quantum gravity may be frozen into large scale one by inflation and the gravitation theory governing the evolution of the universe should take the large scale Lorentz violation into account at the cosmic scale. The induced torsion can contribute an additional energy-momentum distribution accompanying the cosmic media source and gives rise to the accelerated expansion of the universe\cite{Shen:2018elj, Han-Yu:2019tmf, Zhai:2019std, Li:2020tqx, Li:2021gab}. The cosmic torsion can influence the propagation path of the neutrino traveling at the cosmic distance. The possible different responce manners of Majorana and Dirac particle to torsion may also distinguish the fermion type of neutrino.

\section{ Tetrad fields and spin connection}

The tetrad fields ${e^\mu}_a(x),a=0,1,2,3$ are defined as

\begin{equation}
{e^\mu }_a{e^\nu }_b{g_{\mu \nu }} = {\eta _{ab}}\;,
\end{equation}
where we use Greek and Latin letters to represent space-time and tangent space index respectively. Then we can define Gamma matrices with space-time index 
\begin{equation}
{\gamma ^\mu }(x) = {e^\mu }_a(x){\gamma ^a}
\end{equation}
and spin connection
\begin{equation}\label{spin-connection}
{\Gamma _{\mu a}}^b = {e_\beta }^b{\Gamma _{\mu \alpha }}^\beta {e^\alpha }_a + {e_\alpha }^b{\partial _\mu }{e^\alpha }_a = {e_\alpha }^b{\nabla _\mu }{e^\alpha }_a\;.
\end{equation}
The connection can be decomposed into Levi-Civita one and contortion\cite{Hehl:1976kj}
\begin{equation}
    {\Gamma _{\mu \nu }}^\lambda  ={ { \tilde\Gamma} _{\mu \nu }}^\lambda  + {K_{\mu \nu }}^\lambda \;,
\end{equation}
where ${ { \tilde\Gamma} _{\mu \nu }}^\lambda$ is the Christoffel symbol, spacetime indexed Levi-Civata connection and ${K_{\mu \nu }}^\lambda$ is the contortion tensor. 
It can be derived from the formula for the spin connection \eqref{spin-connection} that  
\begin{equation}\label{Gamantismt}
{\Gamma _{\mu ab}} = {e_{\alpha b}}{\nabla _\mu }{e^\alpha }_a = {\nabla _\mu }\left({e_{\alpha b}}{e^\alpha }_a\right) - {e^\alpha }_a{\nabla _\mu }{e_{\alpha b}} =  - {e^\alpha }_a{\nabla _\mu }{e_{\alpha b}} =  - {\Gamma _{\mu ba}}\;,
\end{equation}
i.e. the last two indices of spin connection are anti-symmetric if metricity condition ${\nabla _\mu }\left({e_{\alpha b}}{e^\alpha }_a\right)=0$ holds. The metric preserving property of spin connection indicates that the spin connection is Lorentzian, i.e. it takes value in the Lorentz Lie algebra. The spin connection defines the Fock-Ivanenko covariant derivative 
\begin{equation}\label{FockIvanenkocd}
	{D_\mu }  = {\partial _\mu }  - \frac{i}{2}{\Gamma _{\mu ab}}{S^{ab}}\;,
\end{equation}
where the second part acts only on the tangent space indices and the Lorentz group generators $S^{ab}$ act on field according to the specific Lorentz representation that the field belongs. The Fock-Ivanenko covariant derivative of a field transforming according to some specific representation of local Lorentz group falls to the tensor product of vector and the representation of the field in consideration so it is the covariant derivative of local Lorentz gauge transformation in this sense.
For example, the covariant derivative of vector fields and spinor fields are
\begin{equation}\label{cdvector}
{D_\mu }{A^d} = {\partial _\mu }{A^d} - \frac{i}{2}{\Gamma _{\mu ab}}{({V^{ab}})_c}^d{A^c} = {\partial _\mu }{A^d} + {\Gamma _{\mu b}}^d{A^b}
\end{equation}
and
\begin{equation}\label{cdspinor}
{D_\mu }\psi  = {\partial _\mu }\psi  - \frac{i}{2}{\Gamma _{\mu ab}}{S^{ab}}\psi 
\end{equation}
respectively,
where
\begin{equation}\label{vectorrep}
{({V^{ab}})_c}^d = i({\eta ^{bd}}{\delta ^a}_c - {\eta ^{ad}}{\delta ^b}_c)
\end{equation}
and
\begin{equation}\label{spinorrep}
{S^{ab}} = \frac{1}{2}{\sigma ^{ab}} = \frac{i}{4}[{\gamma ^a},{\gamma ^b}]	
\end{equation} 
are the vector representation and spinor representation of generators respectively. For scalar field $\phi$, the generators in the  Fock-Ivanenko covariant derivative \eqref{FockIvanenkocd} are 
\begin{equation}\label{scalarrep}
	{S^{ab}} =0\;.
\end{equation}
The spacetime tensor field transforms in the way of the same type of tensor under local Lorentz transformation. Hence the covariant derivative \eqref{FockIvanenkocd} acts on spacetime indices and tangent space ones separately according to different Lorentz representations that the index belongs, e.g.
\begin{equation}\label{cdofdphi}
{D_\mu }({\partial _\nu }\psi ) = {\partial _\mu }({\partial _\nu }\psi ) - \frac{i}{2}{\Gamma _{\mu ab}}{S^{ab}}{\partial _\nu }\psi  + \frac{i}{2}{\Gamma _{\mu ab}}{({V^{ab}})_\nu }^\lambda {\partial _\lambda }\psi = {\nabla _\mu }({\partial _\nu }\psi ) - \frac{i}{2}{\Gamma _{\mu ab}}{S^{ab}}{\partial _\nu }\psi \;.
\end{equation}
Utilizing this property of the covariant derivative, one can prove
\begin{equation}\label{fieldstrength}
[{D_\mu },{D_\nu }] =  - {T_{\mu \nu }}^\lambda {D_\lambda } - \frac{i}{4}{R_{\mu \nu ab}}{\sigma ^{ab}}\;,
\end{equation}
where ${T_{\mu \nu }}^\lambda$ is the torsion tensor and $R_{\mu \nu ab}$ is the curvature tensor respectively. The torsion and curvature hence can be viewed as the gauge field strength of local Lorentz group and local translation group respectively\cite{Hehl:1976kj}.
Detailed derivation reveals 
\begin{equation}
\begin{aligned}
\left[{D_\mu },{\gamma ^\nu }\right]&= ({\nabla _\mu }{e^\nu }_c){\gamma ^c} - \frac{i}{2}{\Gamma _{\mu ab}}{e^\nu }_c[{S^{ab}},{\gamma ^c}]\\
 &= ({\nabla _\mu }{e^\nu }_c){\gamma ^c} + {\Gamma _{\mu ab}}{e^{\nu a}}{\gamma ^b}\\
 &= {e^{\nu a}}\left[ {({e_{\lambda a}}{\nabla _\mu }{e^\lambda }_c){\gamma ^c} + {\Gamma _{\mu ab}}{\gamma ^b}} \right]\\
 &= {e^{\nu a}}\left[ {{\Gamma _{\mu ba}} + {\Gamma _{\mu ab}}} \right]{\gamma ^b}\\
 &= 0\;.
\end{aligned}
\end{equation}
So for a spinor field $\psi$, we have 
\begin{equation}\label{cdpsi}
{D_\mu }{\gamma ^\nu }\psi  = {\gamma ^\nu }{D_\mu }\psi	
\end{equation}
or more generally
\begin{equation}\label{cdgama}
{D_\mu }{\gamma ^\nu } = 0\;.	
\end{equation}

\section{Fermion scattering cross section by gravitational field}
\subsection{Fermion Lagrangian in curved spacetime background}
To investigate the Fermion scattering cross section by gravitational field in quantum field theory scheme, one needs to write down the Lagrangian of Fermion in the curved spacetime background. Replacing the partial derivative with covariant derivative in the Lagrangian of fermion in flat space-time,
\begin{equation}
\mathcal{L}_0 = \bar \psi ( {\partial}  - m)\psi\;,
\end{equation}
one can get the Lagrangian in curved space-time,
\begin{equation}\label{diracactcst}
\mathcal{L} = \bar \psi (i{\gamma ^\mu }{D_\mu } - m)\psi\;.
\end{equation}
Observing that the Lagrangian \eqref{diracactcst} is not self-adjoint, 
\begin{equation}
\begin{aligned}
{\mathcal{L}^ * } &= {\mathcal{L}^\dagger  }\\
 &= {\left[ {\bar \psi \left( {i{\gamma ^\mu }(x){\partial _\mu } + \frac{1}{4}{\gamma ^\mu }(x){\Gamma _\mu } - m} \right)\psi } \right]^ \dagger }\\
 &=  - {\partial _\mu }\bar \psi i{\gamma ^\mu }(x)\psi  + \frac{1}{4}{\psi ^ + }{\gamma ^0}{\Gamma _\mu }{\gamma ^0}{\gamma ^0}{\gamma ^\mu }(x)\psi  - m\bar \psi \psi \\
 &=  - {\nabla _\mu }\left( {\bar \psi i{\gamma ^\mu }(x)\psi } \right) + \bar \psi i{\nabla _\mu }{\gamma ^\mu }(x)\psi  + \bar \psi i{\gamma ^\mu }(x){\partial _\mu }\psi  + \frac{1}{4}\bar \psi {\Gamma _\mu }{\gamma ^\mu }(x)\psi  - m\bar \psi \psi \\
 &= \mathcal{L} - {\nabla _\mu }\left( {\bar \psi i{\gamma ^\mu }(x)\psi } \right) + \bar \psi i[{D_\mu },{\gamma ^\mu }(x)]\psi \\
 &= \mathcal{L} - {\nabla _\mu }\left( {\bar \psi i{\gamma ^\mu }(x)\psi } \right)\;,
\end{aligned}
\end{equation}
where ${\Gamma _\mu } = {\Gamma _{\mu ab}}{\sigma ^{ab}}$. It differs from its real part only a total divergence term, ${\nabla _\mu }\left( {\bar \psi i{\gamma ^\mu }(x)\psi } \right)$. 
$\mathcal{L}$ can be re-written as 
\begin{equation}\label{rewriteL}
\mathcal{L}=\frac{1}{2}(\mathcal{L}+\mathcal{L}^*)+\frac{1}{2}{\nabla _\mu }\left( {\bar \psi i{\gamma ^\mu }(x)\psi } \right)	
\end{equation}
and is indeed a well defined Lagrangian. The corresponding action is
\begin{equation}\label{actcst}
S = \int {{d^4}x\sqrt { - g} \mathcal{L}}	\;.
\end{equation}

We will concentrate on the weak field limit of the static spherically symmetric gravitational field background. To take account the weak field limit, we will consider the metric deviation
\begin{equation}\label{mrcdvt}
{h_{\mu \nu }} = {g_{\mu \nu }} - {\eta _{\mu \nu }}	
\end{equation}
and spin connection ${\Gamma _{\mu ab}}$ as infinitesimal quantities. Up to the first order approximation,  we have 
\begin{equation}\label{exp-g}
{ {\sqrt { - g} } \cong 1- \frac{1}{2}{h^\mu }_\mu }
\end{equation}
and
\begin{equation}\label{exptetrad}
{e^\mu }_a \cong {\delta ^\mu }_a - \frac{1}{2}{h^\mu }_a + {\omega ^\mu }_a\;,
\end{equation}
where ${\omega ^\mu }_a$, satisfying ${\omega _{ab}} =  - {\omega _{ba}}$, are the pure inertial part of tetrad fields with six degrees of freedom generated by local Lorentz transformation. One can always remove the inertial ${\omega ^\mu }_a$ by choosing the appropriate local tetrad fields without losing generality, i.e. choosing the preferred tetrad fields satisfying ${\omega ^\mu }_a=0$.
The spacetime indices and the tangent space one play the same role for the covariant quantities in the first order approximation, e.g. ${h^\mu }_a = {e^\mu }_b{h^b}_a \cong {\delta ^\mu }_b{h^b}_a$.
Up to the first order approximation, the action \eqref{actcst} can be expanded as
\begin{equation}\label{actexp}
S = \int {{d^4}x\bar \psi \left( {i\not \partial  - m - \frac{1}{2}{\gamma ^a}{h^\mu }_a{P_\mu } + \frac{1}{4}{\gamma ^a}{\Gamma _a} - \frac{1}{2}{h^\mu }_\mu (i\not \partial  - m)} \right)\psi } \;.
\end{equation}
So the interaction part between fermion and background gravitational field can be read from formula \eqref{actexp},
\begin{equation}\label{intact}
{S_{int}} = \int {{d^4}x\bar \psi \left( { - \frac{1}{2}{\gamma ^a}{h^\mu }_a{P_\mu } + \frac{1}{4}{\gamma ^a}{\Gamma _a} - \frac{1}{2}{h^\mu }_\mu (i\not \partial  - m)} \right)\psi } \;.
\end{equation}
Take the assumption that the spacetime is asymptotic Minkowskian and the initial and final states are both free fermion state in the flat spacetime, the last term $- \frac{1}{2}{h^\mu }_\mu \bar \psi(i\not \partial  - m)\psi $ of \eqref{intact} does not contribute to the scattering amplitude actually.

\subsection{The Scattering of Dirac Fermion }
The plane wave expansion of field operator for Dirac fermion is
\begin{equation}\label{diracfldopt}
\psi  = \sum\limits_s {\int {{d^3}p} } \frac{1}{{\sqrt {EV} }}\left[ {{a_s}(p){u_s}(p){e^{ - ipx}} + {b^ \dagger }_s(p){v_s}(p){e^{ipx}}} \right]\;,
\end{equation}
where the normalization condition is set to
\begin{equation}\label{nmlcdt1}
{{\bar u}_s}(p){u_r}(p) = m{\delta _{sr}}
\end{equation}
and
\begin{equation}\label{nmlcdt2}
{{\bar v}_s}(p){v_r}(p) =  - m{\delta _{sr}}\;.	
\end{equation}
To the first order approximation of spherically symmetric gravitational field, the Schwarzschild metric can be approximated by
\begin{equation}\label{schhmm}
{h_{\mu \nu }} = 2\phi (r){\delta ^\mu }_\nu
\end{equation}
and the spin connection takes the form\cite{2005Neutrino, Lambiase:2005gt}
\begin{equation}\label{schgmmu}
{\Gamma _\mu } =  - \frac{1}{2}{\phi _{,\nu }}{\sigma ^{\mu \nu }}\;,
\end{equation}
where $\phi (r) =  - GM/r$ and the imbalanced indices is just a matter of convention.
Denoting the momentum of the particle in the initial state $\left| i \right\rangle $ and final state $\left| f \right\rangle $ with $\vec k$ and $\vec k'$, the corresponding spin state with $r$ and $s$  respectively, the amplitude for the transition $\left| i \right\rangle \to \left| f \right\rangle$ is given by
\begin{equation}\label{smtrx}
	\left\langle f \right|S \left| i \right\rangle=\left\langle f \right|Texp(-i\int d^4x \mathcal {H}_I(x)) \left| i \right\rangle\;,
\end{equation} 
where $S$ and $\mathcal {H}_I(x)$ are the $S$-matrix and the interacting Hamiltonian respectively. For simplicity, we take the lowest order non-trivial expansion of the scattering amplitude, i.e. the first order perturbation, into consideration, 
\begin{equation}\label{sctamplt}
\begin{aligned}
M_{Dirac} &= i\left\langle f \right|{S_{{\mathop{\rm int}} }}\left| i \right\rangle \\
 &= i\left\langle f \right|\int {{d^4}x\bar \psi \left( { - \frac{1}{2}{\gamma ^a}{h^\mu }_a{P_\mu } + \frac{1}{4}{\gamma ^a}{\Gamma _a}} \right)\psi } \left| i \right\rangle \\
 &= \frac{{2\pi i}}{{VE}}\delta (E - E'){{\bar u}_s}(k')\left( {  \frac{i}{8}{\gamma ^a}{q_\nu \phi }(q){\sigma ^{a\nu }} - \phi (q){\eta _{\mu a} }{\gamma ^a}{k_\mu }} \right){u_r}(k)\\
 &=  - \frac{{2\pi i}}{{VE}}\delta (E - E')\phi (q){{\bar u}_s}(k')\not \tilde  k{u_r}(k)\;,
\end{aligned}
\end{equation}
where $\vec q = \vec k' - \vec k$, the transferred momentum, $|\vec q| = 2|\vec k|\sin (\theta /2)$ if the scattering angle is $\theta$, $\phi (q) = {4\pi GM}/{|\vec q{|^2}}$, the three dimensional Fourier component of $\phi (r)$, $q^0=0$ for the energy conservation and ${{\tilde{k}}^{\mu }}=(E,-\vec{k})={{k}_{\mu }}$.
To derive the scattering cross section for unpolarized particles, one needs to make summation over final spin states and to take average on the initial spin states. The standard derivation in quantum field theory(e.g. \cite{Nair:2005iw}) gives
\begin{equation}\label{diffcrsct}
\frac{{d\sigma }}{{d\Omega }} = \frac{{{G^2}{M^2}}}{{4{v^4}{{\sin }^4}(\theta /2)}}\left[ {{{\left( {1 + {v^2}} \right)}^2} - {v^2}\left( {3 + {v^2}} \right){{\sin }^2}\frac{\theta }{2}} \right]\;,
\end{equation}
where $v$ is velocity of particle($c=1$). Obviously, the cross section \eqref{diffcrsct} only depends on the velocity of the scattered particle.
The non-relativistic limit of the cross section \eqref{diffcrsct} gives 
\begin{equation}\label{nonrltvtdcs}
\mathop {\lim }\limits_{v \to 0} \frac{{d\sigma }}{{d\Omega }} = \frac{{{G^2}{M^2}{m^2}}}{{16{E_k}^2{{\sin }^4}(\theta /2)}}\;,
\end{equation}
where ${E_k} = m{v^2}/2$ is non-relativistic kinetic energy. Formula \eqref{nonrltvtdcs} is just the familiar form of Rutherford scattering cross section. Considering the small angle scattering($\sin \theta \cong \theta$), the ultra-relativistic limit of the cross section \eqref{diffcrsct} reduces to
\begin{equation}\label{relcrosssection}
\underset{v\to 1}{\mathop{\lim }}\,\frac{d\sigma }{d\Omega }=\frac{16{{G}^{2}}{{M}^{2}}}{{{\theta }^{4}}}\;,
\end{equation}
which gives the familiar relationship between light deflection angle and aiming distance $b$,
\begin{equation}\label{aimdis}
\theta  = \frac{{4GM}}{b}\;.
\end{equation}
The fact that the cross section \eqref{diffcrsct} coincides with result in classical theory in both the relativistic and the non-relativistic limit indicates that the lowest order expansion of the scattering amplitude \eqref{sctamplt} can describe the most important feathers of the fermion scattering by gravitational field.

\subsection{The Scattering of Majorana Fermion}
The plane wave expansion of the field operator for Majorana fermion is\cite{2004QuantumFT}
\begin{equation}\label{fldoptmjnfm}
{\psi _M} = \sum\limits_s {\int {{d^3}k} } \frac{1}{{\sqrt {EV} }}\left[ {{a_s}(k){u_s}(k){e^{ - ikx}} + {a^ \dagger }_s(k){v_s}(k){e^{ikx}}} \right]\;.
\end{equation}
The spinor wave function ${v_s}(k)$ and $u_s(k)$ are the charge conjugation of each other,
\begin{equation}\label{spnwf}
{v_s}(k) = {u^c}_s(k) = C{\bar u^T}_s(k)\;,
\end{equation}
where $C = i{\gamma ^0}{\gamma ^2}$ is the charge conjugation transformation. The Majorana field operator is self charge conjugated,
\begin{equation}\label{mjncgcjg}
{\psi _M} = C{\bar \psi _M}^T = {\psi _M}^c\;,
\end{equation}
which means Majorana fermion is the antiparticle of itself. 
There is almost no difference between the actions of Dirac and Majoranan fermion except for a factor one half,
\begin{equation}\label{actmjnfm}
	\mathcal{L}_M = \frac{1}{2}{\bar \psi _M}(i{\gamma ^\mu }{D_\mu } - m){\psi _M}\;,
\end{equation}
for Dirac field describes two kinds of particle while only one kind of particle does Majorana field. Similar derivation of formula \eqref{intact} gives the interaction action between Majorana fermion and background gravitational field,
\begin{equation}\label{intactmjn}
{S_{int,M}} = \frac{1}{2}\int {{d^4}x{{\bar \psi }_M}\left( { - \frac{1}{2}{\gamma ^a}{h^\mu }_a{P_\mu } + \frac{1}{4}{\gamma ^a}{\Gamma _a} - \frac{1}{2}{h^\mu }_\mu (i\not \partial  - m)} \right){\psi _M}} \;.
\end{equation}
By the same consideration in deriving scattering amplitude for Dirac fermion \eqref{sctamplt}, the scattering amplitude of Majorana fermion by spherically symmetric gravitational field in the lowest order  expansion of the interaction can be derived 
\begin{equation}\label{sctamplmjrnfm}
\begin{aligned}
& {{M}_{Majorana}}=\frac{\pi i}{VE}\delta \left( E-{{E}^{\prime }} \right)\left[ {{{\bar{u}}}_{s}}\left( {{k}^{\prime }} \right)\left( \frac{1}{4}{{\gamma }^{a}}{{\Gamma }_{a}}(q)-\frac{1}{2}{{h}^{\mu }}_{a}(q){{\gamma }^{a}}{{k}_{\mu }} \right){{u}_{r}}(k) \right. \\ 
& \left. -{{{\bar{v}}}_{r}}(k)\left( \frac{1}{4}{{\gamma }^{a}}{{\Gamma }_{a}}(q)+\frac{1}{2}h_{a}^{\mu }(q){{\gamma }^{a}}k_{\mu }^{\prime } \right){{v}_{s}}\left( {{k}^{\prime }} \right) \right] \;.\\ 
\end{aligned}
\end{equation}
where ${{\Gamma _a}(q)}$ and ${{h^\mu }_a(q)}$ are the three dimensional Fourier components of the corresponding ${{\Gamma _a}(x)}$ and ${{h^\mu }_a(x)}$. The first part of $M_{Majorana}$ contains $u$ spinor only and is the same as in $M_{Dirac}$ while the second part contains $v$ spinor only. Utilizing the charge conjugation relation \eqref{mjncgcjg} of spinors, a similar derivation of \eqref{sctamplt} gives 
\begin{equation}\label{smplmjnsctamp}
\begin{aligned}
&{{\bar v}_r}(k)\left( {\frac{1}{4}{\gamma ^a}{\Gamma _a}(q) + \frac{1}{2}{h^\mu }_a(q){\gamma ^a}{{k'}_\mu }} \right){v_s}(k')\\
 &= \phi (q){{{\bar{v}}}_{r}}(k)\left( \frac{1}{8}\not{q}+{\tilde{\not{k}}}' \right){{v}_{s}}({k}') \\
 &= \phi (q) {  {{\bar v}_r}(k)\tilde \not k'{v_s}(k')} \\
 & =-\phi (q){{u}_{r}}^{T}(k){{C}^{-1}}{\tilde{\not{k}}}'C{{{\bar{u}}}^{T}}_{s}({k}') \\ 
 & =\phi (q){{u}_{r}}^{T}(k){{({\tilde{\not{k}}}')}^{T}}{{{\bar{u}}}^{T}}_{s}({k}') \\ 
 &= \phi (q){{\bar u}_s}(k')\left( { + \tilde \not k'} \right){u_r}(k)\\
 &= \phi (q){{\bar u}_s}(k')\left( {\tilde \not k - \not q} \right){u_r}(k)\\
 &= \phi (q){{\bar u}_s}(k')\tilde \not k{u_r}(k)\;,
\end{aligned}
\end{equation}
where ${\tilde  k}^{\prime \mu} = k'_\mu$. It leads to $M_{Majorana}=M_{Dirac}$, which means two kinds of fermions will have the same scattering behavior in spherically symmetric gravitational field in the lowest order expansion of the interaction. 

The treatment of Dirac and Majorana fermion scattering \eqref{sctamplt} and \eqref{sctamplmjrnfm} can apply to the neutrino scattering case with the insertion of chiral projection operator ${P_L} = (1 - {\gamma _5})/2$ for neutrino is only left handed with positive left hand helicity. Neglecting the small mass of neutrino, the scattering amplitude of Dirac neutrino is 
\begin{equation}\label{sctampldirntrn}
{M_{Dirac}^N} =  - \frac{{2\pi i}}{{VE}}\delta (E - E')\phi (q){\bar u_s}(k')\tilde \not k{P_L}{u_r}(k)\;.
\end{equation}
For Majorana neutrino, the scattering amplitude is 
\begin{equation}\label{sctampmjnntrn}
\begin{aligned}
 & M_{Majorana}^{N}=\frac{\pi i}{VE}\delta (E-{E}')\left[ {{{\bar{u}}}_{s}}({k}')\left( \frac{1}{4}{{\gamma }^{a}}{{\Gamma }_{a}}(q)-\frac{1}{2}{{h}^{\mu }}_{a}(q){{\gamma }^{a}}{{k}_{\mu }} \right){{P}_{L}}{{u}_{r}}(k) \right. \\ 
& \left. -{{{\bar{v}}}_{r}}(k)\left( \frac{1}{4}{{\gamma }^{a}}{{\Gamma }_{a}}(q)+\frac{1}{2}{{h}^{\mu }}_{a}(q){{\gamma }^{a}}{{{{k}'}}_{\mu }} \right){{P}_{R}}{{v}_{s}}({k}') \right]\;. \\ 
\end{aligned}
\end{equation}
However, a similar derivation as in \eqref{smplmjnsctamp} shows that the relation $M_{Majorana}^N=M_{Dirac}^N\equiv M_N$ 
still holds. 

Formula \eqref{sctampldirntrn} shows that the scattering amplitude is non-zero only when the helicities of the initial and final states of neutrino are the same left handed one. For neutrino scattering, the summation over final spin states and average over the initial states are trivial,
\begin{equation}\label{polsummntrin}
|{{M}_{N}}{{|}^{2}}={{\sum\limits_{spins}{|{{{{M}}}_{N}}|}}^{2}}	\;.
\end{equation}
The spinor calculation gives
\begin{equation}\label{sctampsquaresummavrage}
\begin{aligned}
	& \sum\limits_{spins}{|{{{\bar{u}}}_{s}}({k}')\tilde{\not{k}}{{P}_{L}}{{u}_{r}}(k)}{{|}^{2}} \\ 
	& =\sum\limits_{spins}{{{{\bar{u}}}_{s}}({k}')\tilde{\not{k}}{{P}_{L}}{{u}_{r}}(k)}{{{\bar{u}}}_{r}}(k)\tilde{\not{k}}{{P}_{L}}{{u}_{s}}({k}') \\ 
	& =\frac{1}{4}Tr\left[ \not{{k}'}\tilde{\not{k}}{{P}_{L}}\not{k}\tilde{\not{k}}{{P}_{L}} \right] \\ 
	& =\frac{1}{8}Tr\left[ \not{{k}'}\tilde{\not{k}}\not{k}\tilde{\not{k}}(1-{{\gamma }_{5}}) \right] \\ 
	& =\frac{1}{2}\left[ 2\tilde{k}\cdot {k}'k\cdot \tilde{k}+i{{\varepsilon }^{\mu \nu \rho \lambda }}{{{{k}'}}_{\mu }}{{{\tilde{k}}}_{\nu }}{{k}_{\rho }}{{{\tilde{k}}}_{\lambda }} \right] \\ 
	& =4{{E}^{4}}(1-{{\sin }^{2}}\frac{\theta }{2})\;. \\ 
\end{aligned}
\end{equation}
The differential cross section can be given by
\begin{equation}\label{crosssectionneutrino}
\frac{d{{\sigma }_{N}}}{d\Omega }=\frac{{{G}^{2}}{{M}^{2}}}{{{\sin }^{4}}(\theta /2)}\left[ 1-{{\sin }^{2}}\frac{\theta }{2} \right]	\;,
\end{equation}
which is just the $v\approx c$ limit in \eqref{diffcrsct}. 

In conclusion, up to the lowest order expansion in the interaction of neutrino and gravity, the scattering behavior of neutrino in the spherically symmetric gravitational field background can not distinguish the type of neutrino from Dirac one to Majorana one.
 
\subsection{The Effect of Torsion on Scattering}
Torsion may be an essential part of gravity from a point of view of either Einstein-Cartan extension of general relativity or the equivalent description of gravity by both curvature and torsion. To get the torsion involved in the scattering of fermion by gravity, one needs a detailed investigation of the interaction between torsion and fermion. The torsion tensor $T^{\mu \nu \rho }$ can be decomposed into three parts according to the irreducible representation of Lorentz group, i.e. the vector torsion $V^\mu$,the axial vector torsion $A^\mu$ and the pure tensor torsion $F^{\mu \nu \rho}$\cite{Aldrovandi:2013wha},
\begin{equation}\label{decmtort}
\begin{aligned}
{V^\mu } &= {T^{\mu \nu }}_\nu \\
{A^\mu } &= \frac{1}{{3!}}{\varepsilon ^{\mu \nu \rho \lambda }}{T_{\nu \rho \lambda }}\\
{F^{\mu \nu \rho }} &= {T^{\mu (\nu \rho )}} + \frac{1}{3}{g^{\mu (\nu }}{V^{\rho )}} - \frac{1}{3}{g^{\nu \rho }}{V^\mu }\;.
\end{aligned}
\end{equation}
The pure tensor torsion does not involve in the coupling between torsion and fermion and the covariant derivative of spinor field can be written as
\begin{equation}\label{cdfemtors}
{\gamma ^\mu }{D_\mu } = \not \partial  - \frac{i}{4}{\gamma ^\mu }{\tilde \Gamma _{\mu ab}}{\sigma ^{ab}} + \frac{1}{2}\eta_1{\gamma ^\mu }{V_\mu } + \frac{{3i}}{4}{\gamma ^\mu }{A_\mu }{\gamma _5}
\end{equation}
where $\eta_1=0$ in the case of minimal coupling of torsion with fermion and $\eta_1\ne 0$ in the non-minimal coupling scheme as well as the framework of equivalent description of gravity by torsion, e.g. $\eta_1=1$ in the framework of teleparallel gravity. For simplicity, we keep $\eta_1$ as te coupling constant of the vector torsion term in the following of the paper to account for the latter two cases. 

The scattering amplitude of Dirac fermions under the presence of torsion is
\begin{equation}\label{sctamptors}
{M^T_{Dirac}} = \frac{{2\pi i}}{{VE}}\delta (E - E'){\bar u_s}(k')\left( {\frac{1}{4}{\gamma ^a}{{\tilde \Gamma }_a}(q) + \frac{i}{2}\eta_1\not V(q) - \frac{3}{4}\not A(q){\gamma _5} - \frac{1}{2}{h^\mu }_a(q){\gamma ^a}{k_\mu }} \right){u_r}(k)\;.
\end{equation}
The vector part contributes a vector current coupling, while the axial vector part contributes an axial vector current coupling. Similar to the Majorana fermion scattering amplitude formula \eqref{sctamplmjrnfm} by gravity , the scattering amplitude of the Majorana fermion under the presence of torsion can be written as
\begin{equation}\label{sctampmajrntors}
\begin{array}{l}
{M^T_{Majorana}} = \frac{{\pi i}}{{VE}}\delta (E - E'){{\bar u}_s}(k')\left( {\frac{1}{4}{\gamma ^a}{{\tilde \Gamma }_a}(q) + \frac{i}{2}\eta_1\not V(q) - \frac{3}{4}\not A(q){\gamma _5} - \frac{1}{2}{h^\mu }_a(q){\gamma ^a}{k_\mu }} \right){u_r}(k)\\
 - \frac{{\pi i}}{{VE}}\delta (E - E'){{\bar v}_r}(k)\left( {\frac{1}{4}{\gamma ^a}{{\tilde \Gamma }_a}(q) + \frac{i}{2}\eta_1\not V(q) - \frac{3}{4}\not A(q){\gamma _5} + \frac{1}{2}{h^\mu }_a(q){\gamma ^a}{{k'}_\mu }} \right){v_s}(k')\;.
\end{array}
\end{equation}
The first term in \eqref{sctampmajrntors} is one half of the amplitude of Dirac fermion scattering \eqref{sctamptors} again just as torsion free case in \eqref{sctamplmjrnfm}. The effect of torsion can be separated from the full expressions \eqref{sctamptors} and \eqref{sctampmajrntors}. As in the torsion free derivation \eqref{smplmjnsctamp}, the second part of torsion terms from the Majorana particle scattering \eqref{sctampmajrntors} can be converted to  
\begin{equation}\label{secdmajrntors}
{\bar v_r}(k)\left( {\frac{i}{2}\eta_1\not V(q) - \frac{3}{4}\not A(q){\gamma _5}} \right){v_s}(k') = {\bar u_s}(k')\left( {\frac{i}{2}\eta_1\not V(q) + \frac{3}{4}\not A(q){\gamma _5}} \right){u_r}(k)\;,
\end{equation}
which reduces the full expression \eqref{sctampmajrntors} to
\begin{equation}\label{majnsctamptorsn}
\begin{array}{l}
{M^T_{Majorana}} = \frac{{\pi i}}{{VE}}\delta (E - E'){{\bar u}_s}(k')\left( {\frac{1}{4}{\gamma ^a}{{\tilde \Gamma }_a}(q) - \frac{1}{2}{h^\mu }_a(q){\gamma ^a}{k_\mu }} \right){u_r}(k)\\
 - \frac{{\pi i}}{{VE}}\delta (E - E'){{\bar v}_r}(k)\left( {\frac{1}{4}{\gamma ^a}{{\tilde \Gamma }_a}(q) + \frac{1}{2}{h^\mu }_a(q){\gamma ^a}{{k'}_\mu }} \right){v_s}(k')\\
 + \frac{{2\pi i}}{{VE}}\delta (E - E'){{\bar u}_s}(k')\left( { - \frac{3}{4}\not A(q){\gamma _5}} \right){u_r}(k)\;.
\end{array}
\end{equation}
The expression \eqref{majnsctamptorsn} reveals that the vector torsion does not play any role in the scattering amplitude of the Majorana fermion, while the axial vector torsion contributes the same term in Majorana case just as in the Dirac fermion case up to the first order of perturbation. The result is independent of the spacetime background actually although it is derived in a weak field approximation of the spherically symmetric background. The conclusion may be a reflection of the fact that the Majorana fermion has neither a $U(1)$ symmetry hence nor a conserved $U(1)$ current which can couple to the vector torsion to contribute to the scattering amplitude. The possible different response to the vector torsion between Majorana and Dirac particle can distinguish the fermion type of neutrino from the scattering behavior by vector torsion. It can also serve as a prediction of the non-minimal coupling between torsion and fermion or a verification of the equivalence between the curvature and the torsion descriptions of gravity.

\section{Summary and Outlook}

The conclusion that Dirac and Majorana fermion are indistinguishable by the torsionless spherically symmetric gravitational field can not be regarded as a counter proof of the conclusion of Papini et.al. in literature \cite{Singh:2006ad} because the Lense-Thirring background considered in \cite{Singh:2006ad} is not spherically symmetric and is generated by the rotating mass with angular momentum. The spin-gravity interaction is generated by the gravitomagenetic frame dragging effect of the rotating source. On the other hand, our conclusion is only based on the first order perturbation of the interaction between gravitational field and the fermion field. Results beyond first order perturbation and in the case of gravitational field without spherical symmetry are expected in the subsequent works. However, the response difference on vector torsion coupling between Dirac in \eqref{sctamptors} and Majorana fields in \eqref{majnsctamptorsn} is expected to hold beyond the lowest order perturbation. 

Torsion can be generated from the polarized distribution of spinor field which may happen in some high density astronomical objects such as neutron star or near the blackhole\cite{Singh:2006ad, Hehl:1976kj}. The possible detection of torsion-spinor interaction can be the scattering behavior of neutrino beam passing by these kind of objects. 

In view point of effective field theory, the connection may be an active degree of freedom, which can be excited at high energy scale\cite{Donoghue:2016vck,Donoghue:2016xnh}. In low energy effective theory of gravity, the torsion free Levi-Civita connection may be a confined or condensed phase of the high energy connection degree of freedom. There remains the possibility that the low energy connection phase deviates from Levi-Civita one at the large scale as in the large scale Lorentz violation scheme\cite{wu2015effective, wu2016sim, yang2017VSR, wei2017E2}. The large scale Lorentz violation approach to the gravitational theory at cosmic scale and cosmology predicts a torsion distribution at the cosmic scale\cite{Shen:2018elj, Han-Yu:2019tmf, Zhai:2019std, Li:2020tqx, Li:2021gab}. The prediction and detection on the behavior of neutrino propagation at cosmic scale can verify the predicted cosmic torsion by the large scale Lorentz violation model. However the technique developed in this paper needs modification and extension to the cosmological torsion case. The geometro-hydrodynamics approach may be employed to the prediction of neutrino propagation path in the FRW background with cosmic torsion distribution rather than the scattering of free neutrino by a localized torsion and curvature distribution\cite{Trukhanova:2018gmh,Trukhanova:2017yyq, Shapiro:2001rz}. 

In the framework of equivalent description of gravity by torsion rather than curvature or both of them, the coupling of vector torsion with fermion field does not vanish\cite{Aldrovandi:2013wha,Arcos:2004ig,Hammond:2002rm}. Its correspondence in the Hilbert——Einstein framework needs to be clarified. The possible relation between different response to the vector torsion by Dirac and Majorana fermion in the frame work of teleparallel gravity and spin-gravity coupling induced different spin-flip matrix element between Dirac and Majorana fermion in a rotating gravitational field background\cite{Singh:2006ad,Nieves:2006tq,Singh:2006fn} needs to be investigated further.

%----------------------------------------

\section*{Acknowledgment}
This work is supported by the National Natural Science Foundation of China, under Grant No. 11775080 and Grant No. 11865016.

%\newpage

\bibliographystyle{utcaps}
\bibliography{References}

\end{document}